\documentclass[twocolumn,aps,floatfix,superscriptaddress,longbibliography]{revtex4-1}
\pdfoutput=1 %This is for the ArXiv submission only
\usepackage{amsmath,amssymb,eucal,graphicx,float,epstopdf}
\usepackage{epsfig}
\usepackage[utf8]{inputenc}

\usepackage[colorlinks=true, urlcolor=blue, anchorcolor=blue, citecolor=blue,filecolor=blue,linkcolor=blue,menucolor=blue]{hyperref}

\usepackage{subfigure}

\newcommand{\ii}{{\rm i}}

\begin{document}
\title{Impurity in a zero-temperature three-dimensional Fermi gas}

\author{P. L. Krapivsky}
\affiliation{Department of Physics, Boston University, Boston, Massachusetts 02215, USA}
\affiliation{Santa Fe Institute, Santa Fe, New Mexico 87501, USA}

\begin{abstract}
We consider an impurity in a sea of zero-temperature fermions uniformly distributed throughout the space. The impurity scatters on fermions. On average, the  momentum of impurity decreases with time as $t^{-1/(d+1)}$ in $d$ dimensions, and the momentum distribution acquires a scaling form in the long time limit. We solve the Lorentz-Boltzmann equation for the scaled momentum distribution of the impurity in three dimensions. The solution is a combination of confluent hypergeometric functions. In two spatial dimensions, the Lorentz-Boltzmann equation is analytically intractable, so we merely extract a few exact predictions about asymptotic behaviors when the scaled momentum of the impurity is small or large. 
\end{abstract}

\maketitle

\section{Introduction}

Describing the properties of an impurity interacting with a bath of fermions, a Fermi polaron, remains an intriguing theoretical challenge. This problem has a long and venerable history \cite{Landau-polaron,Landau-Pekar,Meyer,Polaron1,Polaron2}. Experimental progress in cold atoms provides new motivation for studying polaronic phenomena \cite{Bruun08,Zwierlein09,Chevy10,Klawunn,Stoof,Kim-Huse-imp,Castin12,Castin14,Doggen,Bruun14,Cui17,Demler18, Zwierlein19,Zvonarev21,Nishimura,Parish23,Chang23}. An extreme case of an impurity immersed in a zero-temperature Fermi gas has become relevant \cite{Lychkovskiy14b,Lychkovskiy15,Ardila21}, and the control of dimensionality \cite{Batchelor13} suggests investigating polaronic phenomena in various dimensions. In one dimension, the impurity exhibits peculiar behaviors \cite{Fisher,Kamenev09,Kamenev12,Lychkovskiy14c,Lychkovskiy14a,Zvonarev14, Zvonarev15,Zvonarev16,Kamenev16,Zvonarev20,Gamayun} including, e.g., momentum oscillations in the presence of an external force \cite{Kamenev09,Kamenev12,Lychkovskiy14c}, a drastic dependence on whether the masses of the impurity and the host fermions are equal or not \cite{Lychkovskiy14a}, and quantum flutter phenomenon \cite{Zvonarev14}. Qualitatively different and typically more robust behaviors emerge in $d\geq 2$ dimensions, particularly in three dimensions \cite{Kim-Huse-imp,Castin12,Castin14}. 

The behavior of the impurity in a zero-temperature Fermi gas is particularly tractable in the physically important three-dimensional case. In Sec.~\ref{sec:3d}, we show that the governing equation for the scaled momentum distribution admits an analytical solution. In two dimensions, the governing equation for the scaled momentum distribution appears analytically intractable, but one still can extract asymptotic behaviors (Sec.~\ref{sec:2d} and Appendix~\ref{ap:origin}). 

In Sec.~\ref{sec:position}, we outline a few challenges for future work, like analyzing the position distribution and more generally the joint position-momentum distribution of the impurity in a zero-temperature Fermi gas. We also discuss the massless impurity which may provide a curious implementation of the Fermi acceleration phenomenon in the zero-temperature Fermi gas.

\section{Three Dimensions}
\label{sec:3d}

The momentum distribution $F(q,t)$ of an impurity in a zero-temperature Fermi gas evolves according to 
\begin{equation}
\label{LB}
\frac{dF(q,t)}{dt} = \int_q^\infty dQ\,Q(Q^2-q^2)F(Q,t)-\frac{2}{15}\,q^4 F(q,t)
\end{equation}
in three dimensions \cite{Kim-Huse-imp}. Major assumptions underlying the applicability of Eq.~\eqref{LB} are the following: (i) the impurity is treated classically;  (ii) the influence of the impurity on an infinite system of fermions is neglected, so the host fermions remain in a zero-temperature Fermi-Dirac distribution; (iii) the energy of impurity is low compared to Fermi energy; (iv) the momentum distribution is spherically symmetric in the long time limit; see Kim and Huse \cite{Kim-Huse-imp} for explanations and justifications of the above assumptions. For instance, when $q$ is low enough, no internal excitations of the polaron are possible. The average momentum decays with time as we see below, so if $q\ll p_F$ initially, it is expected to be satisfied throughout the evolution thereby supporting (iii).  If $F({\bf q}, 0)$ is anisotropic, $F({\bf q}, t)$ quickly becomes isotropic. We want to understand the asymptotic behavior, so (iv) is valid in the interesting regime. We also tacitly assume that the effective mass of the impurity is comparable with the mass of fermions. (Different behaviors may occur for the massless impurity as we mention in Sec.~\ref{sec:position}.)

It is convenient to measure the momentum of the impurity $q$ in units of the Fermi momentum $p_F$. In Eq.~\eqref{LB}, we set to unity an amplitude in the gain term on the right-hand side. In dimensionful variables, the amplitude involves the (effective) mass of the impurity, the Planck constant, the scattering length, etc. \cite{Kim-Huse-imp}. We absorbed the amplitude into the time variable. An amplitude in the loss term is then fixed by normalization 
\begin{equation}
\label{norm}
\int d{\bf q}\, F({\bf q}, t)  = 4\pi \int_0^\infty dq\, q^2 F(q,t) = 1
\end{equation}
Indeed,
\begin{eqnarray*}
&& \int_0^\infty dq\, q^2\int_q^\infty dQ\,Q(Q^2-q^2)F(Q,t) \\
&&= \int_0^\infty dQ\,QF(Q,t) \int_0^Q dq\, q^2(Q^2-q^2) \\
&&= \frac{2}{15}\int_0^\infty dQ\,Q^6 F(Q,t) 
\end{eqnarray*}
The form of \eqref{norm} accounts that the momentum distribution is spherically symmetric.  As we have already asserted, Eq.~\eqref{LB} is applicable when $q\ll 1$.

Equations resembling \eqref{LB} are known as linear Boltzmann or Lorentz-Boltzmann equations. In the realm of the Lorentz gas \cite{Lorentz1905-I,Hauge,Lorentz}, the impurity is effectively massless, and only the direction of velocity changes in elastic collisions of the impurity with scatters. The velocity distribution satisfies a simple Lorentz-Boltzmann equation. The joint position-velocity distribution of the impurity also satisfies a solvable Lorentz-Boltzmann equation \cite{Hauge}.) The linearity of Lorentz-Boltzmann equations makes them significantly more tractable than non-linear Boltzmann equations \cite{fluid,KRB,Beijeren,kremer10}. 

Our chief interest is the large-time behavior. In this situation, the momentum distribution approaches a scaling form, viz.
\begin{equation}
\label{scaling}
F(q,t) = t^{3/4} g(s), \quad s = t^{1/4} q
\end{equation}
when $t\to\infty$ and $q\to 0$ with scaled momentum $s=t^{1/4} q$ kept finite. In this scaling limit, Eq.~\eqref{LB} reduces \cite{Kim-Huse-imp} to the integro-differential equation 
\begin{eqnarray}
\label{LB:scaled}
\left(\frac{3}{4}+\frac{s}{4}\,\frac{d}{ds}\right)g(s) &=& - \frac{2}{15}\,s^4 g(s) \nonumber \\
&+& \int_s^\infty d\sigma\,\sigma\big(\sigma^2-s^2\big)g(\sigma)
\end{eqnarray}
for the scaled momentum distribution. 

It proves convenient to recast Eq.~\eqref{LB:scaled} to a differential equation. Differentiating \eqref{LB:scaled} yields
\begin{equation}
\label{gG:eq}
g'+\tfrac{1}{4} sg''+\tfrac{2}{15} (s^4 g)' = -2s G(s)
\end{equation}
Here we shortly write $(\cdot)' = d (\cdot)/ds$ and use the auxiliary moment distribution function 
\begin{equation}
\label{G:def}
G(s) = \int_s^\infty d\sigma\,\sigma g(\sigma)
\end{equation}
Re-writing \eqref{gG:eq} in terms of $G$ we arrive at a linear ordinary differential equation 
\begin{equation}
\label{G:eq}
s^2 G'''+2s\big(1+\tfrac{4}{15}s^4\big)G''=2\big(1-\tfrac{4}{5}s^4\big)G' +8s^3 G
\end{equation}
This equation admits a remarkably simple solution 
\begin{equation}
\label{G:sol}
G(s) = AF\big[-\tfrac{3}{4}; \tfrac{1}{2}; -\tfrac{2}{15}s^4\big]-s^2  BF\big[-\tfrac{1}{4}; \tfrac{3}{2}; -\tfrac{2}{15}s^4\big]
\end{equation}
Here $F[a; b; x]$ denotes a confluent hypergeometric function \cite{Knuth} with parameters $a$ and $b$. 

The general solution of the third order linear ordinary differential equation \eqref{G:eq} is a combination of three linearly independent solutions and only two appear in Eq.~\eqref{G:sol}. The general solution of \eqref{G:eq} is given by \eqref{G:sol} plus
\begin{equation}
C\big(45s^{-1}+32s^3\big)
\end{equation}
The $s^{-1}$ divergence at the origin and the $s^3$ divergence at infinity are physically unacceptable, e.g., the normalization requirement is violated [cf. with \eqref{norm:3}]. Therefore the amplitude must vanish, $C=0$. 

Hence the the auxiliary moment distribution is given by \eqref{G:sol}. We should also determine $g(s)$. Combining \eqref{G:def} and \eqref{G:sol}, and recalling the identity \cite{Knuth} 
\begin{equation}
\frac{d}{dx}\,F[a; b; x] = \frac{a}{b}\,F[1+a; 1+b; x]
\end{equation}
we deduce the scaled momentum distribution of an impurity in a zero-temperature Fermi gas in three dimensions
\begin{eqnarray}
\label{g:sol}
g(s) &=& 2B\,F\big[-\tfrac{1}{4}; \tfrac{3}{2}; -\tfrac{2}{15}s^4\big] +B\, \tfrac{4}{45}s^4  F\big[\tfrac{3}{4}; \tfrac{5}{2}; -\tfrac{2}{15}s^4\big]\nonumber \\
&-&A\,\tfrac{4}{5}s^2 F\big[\tfrac{1}{4}; \tfrac{3}{2}; -\tfrac{2}{15}s^4\big] 
\end{eqnarray}

To determine the amplitudes in \eqref{G:sol} and \eqref{g:sol} we require the scaled momentum distribution vanish when $s\to\infty$:
\begin{equation}
\label{Gg:inf}
G(\infty)=g(\infty)=0
\end{equation}
Also, the normalization \eqref{norm} must be obeyed.  In terms of the scaled momentum distribution, Eq.~\eqref{norm} becomes 
\begin{equation}
\label{norm:3}
4\pi \int_0^\infty ds\, s^2 g(s) = 4\pi \int_0^\infty ds\, G(s) = 1
\end{equation}
Using Eqs.~\eqref{Gg:inf}--\eqref{norm:3} we fix the amplitudes 
\begin{equation}
\label{AB}
\begin{split}
A& = \frac{\Gamma(1/4)}{5^{1/4}\,(3/2)^{5/4}\,\pi^{5/2}}= 0.083\,493\ldots \\
B& = \frac{5^{1/4}}{6^{3/4}\,\pi^{3/2}\,\Gamma(9/4)} = 0.061\,826\ldots
\end{split}
\end{equation}

\begin{figure}[ht]
\begin{center}
\includegraphics[width=0.45\textwidth]{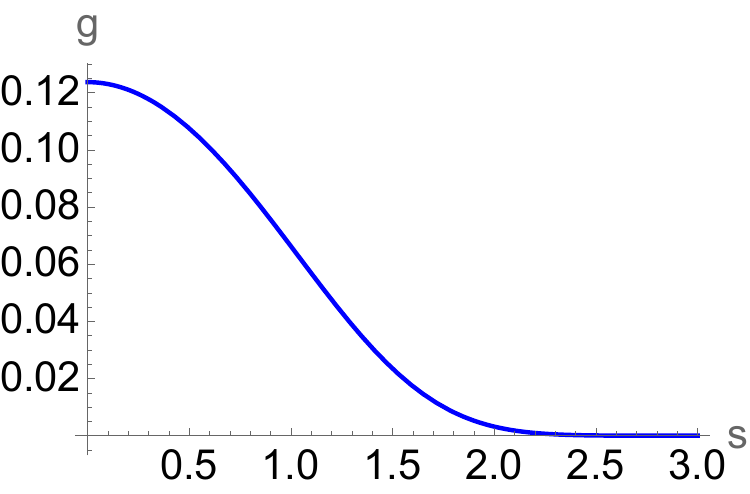}
\caption{The plot of the scaled moment distribution function, Eq.~\eqref{g:sol}, with amplitudes given by \eqref{AB}.}
\label{Fig:gs-norm}
\end{center}
\end{figure}

The asymptotic decay of the scaled momentum distribution is very sharp (see Fig.~\ref{Fig:gs-norm}). The leading asymptotic can be extracted from Eq.~\eqref{G:eq} using the WKB approach \cite{Bender}, viz., seeking the solution in the form $G=e^{-S}$ with rapidly increasing $S$. The dominant exponential decay is $G(s)\propto \mathcal{E}(s)=e^{-2s^4/15}$. A more accurate WKB treatment gives the leading algebraic pre-factor, $G\sim s^{-5}\mathcal{E}(s)$, which in conjunction with \eqref{G:def} yields 
\begin{equation}
\label{g3:asymp}
g\sim s^{-3} \exp\!\big[-\tfrac{2}{15}s^4\big]
\end{equation}

The scaled momentum distribution \eqref{g:sol} is maximal at the origin. Near the origin
\begin{equation}
\label{g3:origin}
g(s)=2B - \tfrac{4}{5}A s^2 +\tfrac{2}{15}B s^4 + \tfrac{4}{225}A s^6 + \ldots
\end{equation}

\section{Two Dimensions}
\label{sec:2d}

The two-dimensional case is also experimentally accessible. The momentum distribution of an impurity in a 2D zero-temperature Fermi gas approaches a scaling form
\begin{equation}
\label{scaling:2d}
F(q,t) = t^{2/3} g(s), \quad s = t^{1/3} q
\end{equation}
The Lorentz-Boltzmann equation for the scaled momentum distribution 
\begin{eqnarray}
\label{LB:scaled-2d}
\left(\frac{2}{3}+\frac{s}{3}\,\frac{d}{ds}\right)g(s) &=&\int_s^\infty d\sigma\,\big(\sigma^2-s^2\big)K\big(\tfrac{s}{\sigma}\big)g(\sigma) \nonumber \\
& - & \frac{5392}{11025}\,s^3 g(s)
\end{eqnarray}
involves the complete elliptic integral of the first kind
\begin{equation}
K(k)=\int_0^\frac{\pi}{2}\frac{d\theta}{\sqrt{1-k^2\sin^2\theta}}
\end{equation}
As in three dimensions, we set to unity the amplitude in the gain term on the right-hand side of Eq.~\eqref{LB:scaled-2d}. The amplitude in the loss term was determined numerically in  \cite{Kim-Huse-imp} and found to be $\approx 0.45$. The precise value appearing in  Eq.~\eqref{LB:scaled-2d} is found from the identity
\begin{equation}
\label{K-1}
\int_0^1 dk\, k\big(1-k^2\big)K(k) = \frac{5392}{11025}
\end{equation}
Indeed, 
\begin{eqnarray}
\label{long-int}
&& \int_0^\infty ds\, s \int_s^\infty d\sigma\,\big(\sigma^2-s^2\big)K\big(\tfrac{s}{\sigma}\big)g(\sigma) \nonumber \\
&&= \int_0^\infty d\sigma\,g(\sigma) \int_0^\sigma ds\, s \big(\sigma^2-s^2\big)K\big(\tfrac{s}{\sigma}\big) \nonumber \\
&&= \int_0^\infty d\sigma\,\sigma^4 g(\sigma) \int_0^1 dk\, k\big(1-k^2\big)K(k) \nonumber \\
&&= \frac{5392}{11025}\int_0^\infty d\sigma\,\sigma^4 g(\sigma)
\end{eqnarray}
ensures that the normalization requirement
\begin{equation}
\label{norm2}
2\pi \int_0^\infty ds\, s g(s) = 1
\end{equation}
is satisfied. 

We have not succeeded in solving Eq.~\eqref{LB:scaled-2d}. Differentiating \eqref{LB:scaled-2d} does not recast it into a differential equation. Some asymptotic behaviors of the scaled distribution $g(s)$ can be established without solving \eqref{LB:scaled-2d}. The asymptotic behavior in the large momentum limit,
\begin{equation}
\label{g2:asymp}
g\sim s^{-2} \exp\!\big[-\tfrac{5392}{11025}s^3\big]
\end{equation}
for $s\to\infty$, can be extracted from \eqref{LB:scaled-2d} using the WKB approach.  

Near the origin
\begin{equation}
\label{g2-exp:origin}
g(s)=g(0) - \frac{9}{64}\,s^2 + \ldots
\end{equation}
The precise value $g(0)$ of the scaled distribution at the origin is unknown, while two first derivatives at the origin are known. The expansion \eqref{g2-exp:origin} resembles the expansion \eqref{g3:origin} in three dimensions. The scaled distribution is maximal at the origin both in $d=2$ and $d=3$. 

The derivation of the expansion \eqref{g2-exp:origin} of the scaled distribution is relegated to Appendix~\ref{ap:origin}. In Appendix~\ref{ap:high}, we outline generalizations to higher dimensions.

\section{Discussion}
\label{sec:position}

The evolution of the momentum ${\bf q}$ is independent of the position ${\bf r}$ of the impurity, so the momentum distribution $F({\bf q}, t)$ satisfies a closed equation. The position of the impurity is coupled with momentum, so the position distribution $\Phi({\bf r}, t)$ does not satisfy a closed equation. Thus one should determine the joint position-momentum distribution $\Pi({\bf r}, {\bf q}, t)$ from which one then extracts the position distribution: $\Phi({\bf r}, t) = \int d{\bf q}\, \Pi({\bf r}, {\bf q}, t)$. Heuristic arguments \cite{Kim-Huse-imp} imply that the length of the last step is typically comparable with total displacement. Therefore the typical distance traveled by the impurity scales as $q t\sim t^{d/(d+1)}$. Computing the position and the joint position-momentum distributions of the impurity in a zero-temperature Fermi gas is the challenge.

The necessity of computing the joint distribution even if one is seeking the position distribution is a rather common phenomenon. For instance, it arose \cite{Luca} for the massless impurity \footnote{The assumption that the impurity is massless is extreme, but the same assumption underlies the classical Lorentz gas where the scatters are immobile \cite{Lorentz1905-I,Hauge,Lorentz,fluid,KRB,Beijeren}. Thus, the impurity does not affect the scatters, which effectively implies that the mass of the impurity is much smaller than the mass of the scatters.} in a monoatomic classical gas at equilibrium at temperature $T>0$. If the massless impurity interacts with host atoms via repulsive $r^{-\lambda}$ potential, the speed distribution approaches a scaling form
\begin{equation}
F(v,t)\sim \tau^{-d/\Lambda}e^{-|v|^\Lambda/\tau},\qquad \Lambda=1+\frac{2(d-1)}{\lambda}
\end{equation}
In the gas of hard spheres ($\lambda=\infty$) of radii $a$, the speed distribution is exponential, $\tau^{-d} e^{-|v|/\tau}$, with dimensionless time $\tau\sim \rho a^{d-1}t\sqrt{T/m}$ where $m$ is the mass and $\rho$ the density of the host atoms. On average, the speed of the massless impurity increases since it more frequently collides with approaching than receding atoms. Thus, the massless impurity in an equilibrium classical gas provides a realization of the Fermi acceleration phenomenon \cite{Fermi}. It would be amusing if the massless Fermi polaron in a zero-temperature Fermi gas exhibited the Fermi acceleration. 

Historically, driven impurities gave birth to the entire subject: Lorentz proposed \cite{Lorentz1905-I} his model as an idealized classical description for electron transport, so in addition to collisions with immobile scatters the impurity is accelerated by an electric field. A constant (on average) drift velocity was originally anticipated \cite{Lorentz1905-I}, yet the lack of dissipation leads to the unbounded growth of the velocity of the massless Lorentz polaron in the Lorentz gas with immobile scatters \cite{Piasecki79,KR97}. In the quantum case, even more intriguing behaviors of the driven impurity with non-vanishing mass have been predicted in one dimension \cite{Kamenev09,Kamenev12,Lychkovskiy14c}. The behavior of the driven impurity in the three-dimensional Fermi gas is probably more robust than in one dimension. 

The perturbation of the host atoms by impurity is usually ignored if we are chiefly interested by the behavior of the impurity. One can certainly do this if the impurity is massless. In the general case, the back reaction on the impurity is asymptotically negligible in $d\geq 2$ dimensions bcause (i) only a finite amount of energy can be transferred to the host atoms in the infinite system, (ii) the perturbation of the host atoms is local and decaying with time, and (iii) repeated collisions are rare when $d\geq 2$. One exception in the classical realm occurs when the host atoms are at zero temperature \cite{AKR}. If host atoms are fermions at zero temperature, the back reaction can still be ignored \cite{Kim-Huse-imp}. 

If the impurity is much more massive than the host atoms, its influence on the host gas is profound. In the classical hard sphere gas, an infinitely heavy particle moving with constant velocity $V$ generates an infinitely strong bow shock if the host spheres are initially at rest. In the quantum case, the sonic speed is finite in the gas of fermions even at zero temperature, $c\sim \hbar \rho^{1/d}/m$. Thus, when an infinitely heavy impurity moves in this host gas in $d\geq 2$ dimensions, a hypersonic bow shock \cite{Landau-FM} is formed only when the Mach number $M=\frac{V}{c}\gg 1$. More subtle behaviors are expected in one dimension \cite{Fisher}.

\appendix
\section{Derivation of \eqref{g2-exp:origin}}
\label{ap:origin}

Specializing \eqref{LB:scaled-2d} to $s=0$ one gets
\begin{equation}
\label{g2:origin}
g(0)=\frac{3\pi}{4} \int_0^\infty d\sigma\,\sigma^2 g(\sigma)
\end{equation}
The integral on the right-hand side, the second moment of the scaled distribution, is unknown. One can try to determine it by multiplying Eq.~\eqref{LB:scaled-2d} by $s^2$ and integrating. This allows one to express the second moment via the fifth moment:
\begin{equation}
\label{2-5}
\int_0^\infty ds\, s^2 g(s) = \frac{21248}{33075}\int_0^\infty  ds\, s^5 g(s)
\end{equation}
The calculation of the double integral is similar to the calculation \eqref{long-int} and uses the identity
\begin{equation}
\label{K-2}
\int_0^1 dk\, k^2\big(1-k^2\big)K(k) = \frac{5456}{19845}
\end{equation}
similar to \eqref{K-1}. One can continue and express the fifth moment via the $8^\text{th}$, etc. The asymptotic behavior of high moments can be extracted with the help of the asymptotic \eqref{g2:asymp}, perhaps allowing to connect $g(0)$ with the amplitude in \eqref{g2:asymp}. This amplitude is unknown and hence omitted in Eq.~\eqref{g2:asymp}. 

Summarizing, we do not know $g(0)$. Surprisingly, one can compute two first derivatives at the origin: $g'(0)=0$ and $g''(0)=-\frac{9}{32}$. Differentiating \eqref{LB:scaled-2d} we obtain
\begin{equation}
\label{g21:origin}
g'+\frac{s}{3}\,g'' =  -\frac{3}{8}\, s + O(s^2)
\end{equation}
In deriving \eqref{g21:origin} we used the normalization condition \eqref{norm2} and identities
\begin{equation}
K(0) = \frac{\pi}{2}\,, \quad \lim_{k\to 0} k^{-1}\,\frac{d K(k)}{dk} = \frac{\pi}{4}
\end{equation}
The announced expansion  \eqref{g2-exp:origin} follows from \eqref{g21:origin}.

\section{High Dimensions}
\label{ap:high}

Generally in $d\geq 2 $ dimensions, the momentum distribution approaches a scaling form
\begin{equation}
\label{scaling:d}
F(q,t) = t^\frac{d}{d+1} g(s), \quad s = t^\frac{1}{d+1} q
\end{equation}
and the scaled momentum distribution obeys
\begin{eqnarray}
\label{LB:scaled-d}
\left(\frac{d}{d+1}+\frac{s}{d+1}\,\frac{d}{ds}\right)g(s) &=&\int_s^\infty d\sigma\,h_d(\sigma,s)g(\sigma)\nonumber \\
& - & C_d\,s^{d+1} g(s)
\end{eqnarray}
We already know $h_2$ and $h_3$. Functions $h_d$ can be computed also for $d>3$. These functions are homogeneous, viz., $h_d(\sigma,s) = \sigma^d H_d(k)$ with $k = s/\sigma$. The normalization requirement gives $C_d = \int_0^1 dk\,k^{d-1} H_d(k)$. 

Applying the WKB approach to  \eqref{LB:scaled-d} yields
\begin{equation}
\label{gd:asymp}
g\sim s^{-d} \exp\!\big[-C_d s^{d+1}\big]
\end{equation}
for $s\gg 1$. This asymptotic is valid for all $d\geq 2$. 

Solving \eqref{LB:scaled-d} is challenging. The physically relevant three-dimensional situation is the most tractable. Simplifications also occur in other odd dimensions. Recall that $h_3$ is a polynomial, while $h_2$ is a transcendental function. The same distinction between odd and even $d$ generally holds \cite{Kim-Huse-imp}. The five-dimensional case is the simplest after $d=3$. The scaled momentum distribution satisfies
\begin{eqnarray}
\label{LB:scaled-5}
\left(\frac{5}{6}+\frac{s}{6}\,\frac{d}{ds}\right)g(s) &=& \int_s^\infty d\sigma\,\sigma
\left(\sigma^4-\tfrac{63}{55}\sigma^2 s^2+\tfrac{8}{55}s^4\right)g(\sigma) \nonumber \\
& - & \tfrac{26}{495}\,s^6 g(s)
\end{eqnarray}
Differentiating \eqref{LB:scaled-5} eliminates the integral on the right-hand side at the cost of introducing two auxiliary functions, $G$ given by \eqref{G:def}, and $H=\int_s^\infty d\sigma\,\sigma^3 g(\sigma)$. Massaging the outcome and using $H' = s^2 G'$ one arrives at a closed linear ordinary differential equation for the auxiliary function $G(s)$:
\begin{eqnarray}
\label{G5:eq}
0 &= &1152 s^5G+\big(1980-912s^6\big)G' +36s\big(13s^6-55\big)G'' \nonumber \\
&+&s^2\big(495+52s^6\big)G'''+165s^3G''''
\end{eqnarray}
The general solution of Eq.~\eqref{G5:eq} remaining finite at the origin is a combination of hypergeometric functions ${}_2F_2$ with four indices:
\begin{eqnarray}
\label{G5:sol}
G(s) &=& C_0\,F\big[-\tfrac{1}{4}-\omega, -\tfrac{1}{4}+\omega; \tfrac{1}{3}, \tfrac{2}{3}; -\tfrac{26}{495}s^6\big] \nonumber \\
&+&C_1 s^2\, F\big[\tfrac{1}{12}-\omega, \tfrac{1}{12}+\omega; \tfrac{2}{3}, \tfrac{4}{3}; -\tfrac{26}{495}s^6\big] \nonumber \\
&+&C_2 s^4\, F\big[\tfrac{5}{12}-\omega, \tfrac{5}{12}+\omega; \tfrac{4}{3}, \tfrac{5}{3}; -\tfrac{26}{495}s^6\big]
\end{eqnarray}
We display all indices and shortly write $F$ instead of ${}_2F_2$; we also use the shorthand notation
\begin{equation}
\omega = \frac{\ii}{12}\sqrt{\frac{11}{13}} 
\end{equation}
Recall that in three dimension, $G(s)$ and $g(s)$ are combination of standard confluent hypergeometric functions ${}_1F_1$, see \eqref{G:sol} and  \eqref{g:sol}.

Fixing the amplitudes $C_0, C_1, C_2$ in the solution \eqref{G5:sol} could be cumbersome. The scaled momentum distribution must vanish when $s\to\infty$, so we have again the boundary condition \eqref{Gg:inf}. The normalization requirement 
\begin{equation}
\label{norm:5}
\frac{8\pi^2}{3} \int_0^\infty ds\, s^4 g(s) = 8\pi^2 \int_0^\infty ds\, s^2 G(s) = 1
\end{equation}
gives another constraint. 

\bibliography{references-gas}

\end{document}